\documentclass[conference]{IEEEtran}
\IEEEoverridecommandlockouts

\usepackage{cite}
\usepackage{amsmath,amssymb,amsfonts}
\usepackage{algorithmic}
\usepackage{graphicx}
\usepackage{textcomp}
\usepackage{xcolor}
\usepackage{fancyhdr}
\usepackage{lipsum}
\usepackage{booktabs}

\def\BibTeX{{\rm B\kern-.05em{\sc i\kern-.025em b}\kern-.08em
    T\kern-.1667em\lower.7ex\hbox{E}\kern-.125emX}}

\begin{document}

\title{Machine-Learned Hamiltonians for Quantum Transport Simulation of Valence Change Memories  
\\
\thanks{We acknowledge funding from the ALMOND project (SNSF Sinergia grant no. 198612) and from the Swiss State Secretariat for Education, Research, and Innovation (SERI) through the SwissChips research project. This work was also supported by the Swiss National Supercomputing Center (CSCS) under project lp16 and lp82.}
}


\author{\IEEEauthorblockN{1\textsuperscript{st} Chen Hao Xia}
\IEEEauthorblockA{\textit{Integrated Systems Laboratory} \\
\textit{ETH Zurich}\\
Zurich, Switzerland \\
chexia@iis.ee.ethz.ch
}
\and

\IEEEauthorblockN{2\textsuperscript{nd} Manasa Kaniselvan}
\IEEEauthorblockA{\textit{Integrated Systems Laboratory} \\
\textit{ETH Zurich}\\
Zurich, Switzerland \\
mkaniselvan@iis.ee.ethz.ch
}

\and
\IEEEauthorblockN{3\textsuperscript{rd} Marko Mladenović}
\IEEEauthorblockA{\textit{Integrated Systems Laboratory} \\
\textit{ETH Zurich}\\
Zurich, Switzerland \\
mmladenovic@iis.ee.ethz.ch
}

\and 
\IEEEauthorblockN{4\textsuperscript{th} Mathieu Luisier}
\IEEEauthorblockA{\textit{Integrated Systems Laboratory} \\
\textit{ETH Zurich}\\
Zurich, Switzerland \\
mluisier@iis.ee.ethz.ch
}
}



\maketitle

\begin{abstract}
The construction of the Hamiltonian matrix \textbf{H} is an essential, yet computationally expensive step in \textit{ab-initio} device simulations based on density-functional theory (DFT). In homogeneous structures, the fact that a unit cell repeats itself along at least one direction can be leveraged to minimize the number of atoms considered and the calculation time. However, such an approach does not lend itself to amorphous or defective materials for which no periodicity exists. In these cases, (much) larger domains containing thousands of atoms might be needed to accurately describe the physics at play, pushing DFT tools to their limit. Here we address this issue by learning and directly predicting the Hamiltonian matrix of large structures through equivariant graph neural networks and so-called augmented partitioning training. We demonstrate the strength of our approach by modeling valence change memory (VCM) cells, achieving a Mean Absolute Error (MAE) of 3.39 to 3.58 meV, as compared to DFT, when predicting the Hamiltonian matrix entries of systems made of $\sim$5,000 atoms. We then replace the DFT-computed Hamiltonian of these VCMs with the predicted one to compute their energy-resolved transmission function with a quantum transport tool. A qualitatively good agreement between both sets of curves is obtained. Our work provides a path forward to overcome the memory and computational limits of DFT, thus enabling the study of large-scale devices beyond current \textit{ab-initio} capabilities. 
\end{abstract}

\section{Introduction} 
As the dimensions of modern-day electronic devices keep decreasing, \textit{ab-initio} calculations are gaining momentum to capture atomistic details and their influence on key performance metrics. Density-functional theory (DFT) lends itself optimally to this task. Besides the energy and wavefunction of atomic systems, it can also return their Hamiltonian matrix \textbf{H}. This quantity can then be passed to a quantum transport (QT) solver to compute the ``current vs. voltage'' characteristics of the targeted device. In structures with atomic disorder such as the amorphous oxides used at the switching layer of resistive random access memories (ReRAM), the construction of \textbf{H} can be tedious. Large unit cells containing thousands of atoms are required to accurately represent ReRAM geometries. As DFT scales with $O(N^{3})$, $N$ being the number of atoms, it becomes prohibitively expensive at these scales, limiting the size of the devices that can be investigated at the \textit{ab-initio} level.

In this work, we present an approach that overcomes this bottleneck by replacing DFT-computed Hamiltonians with machine-learned (ML) ones. We demonstrate this approach by simulating a valence change memory (VCM) cell, a ReRAM type with applications in neuromorphic computing \cite{xiao2024recent}. The VCM of interest is made of a TiN- HfO$_2$- Ti/TiN stack and 5,268 atoms. Its conductance can be modulated by applying an external voltage, which leads to the generation/recombination of oxygen vacancies at the Ti-HfO$_2$ interface and the formation/dissolution of conductive filaments. From a modeling perspective, the underlying structural evolution can be tracked with, for example, kinetic Monte Carlo (KMC) \cite{Kaniselvan2023} or molecular dynamics (MD) \cite{Urquiza2021} simulations. To compute the corresponding ``I-V'' characteristics, the Hamiltonian matrix of samples extracted at regular time intervals should be produced with DFT and passed as input to a QT solver. Bypassing the DFT step with ML requires a model that can learn complex features, e.g., non-regular lattices, the influence of oxygen vacancies, and the interaction between atoms of different types. For that purpose, we adopt an equivariant graph neural network (EGNN) \cite{TFNs} and show that, if it is trained on very few VCM configurations, it can accurately predict the Hamiltonian of large structures with unseen vacancy distributions and filament morphologies. The electrical current calculated with the ML Hamiltonian matrices are in good qualitative agreement with DFT results.

\begin{figure}[!t]
    \centering
    \includegraphics[width=1\linewidth]{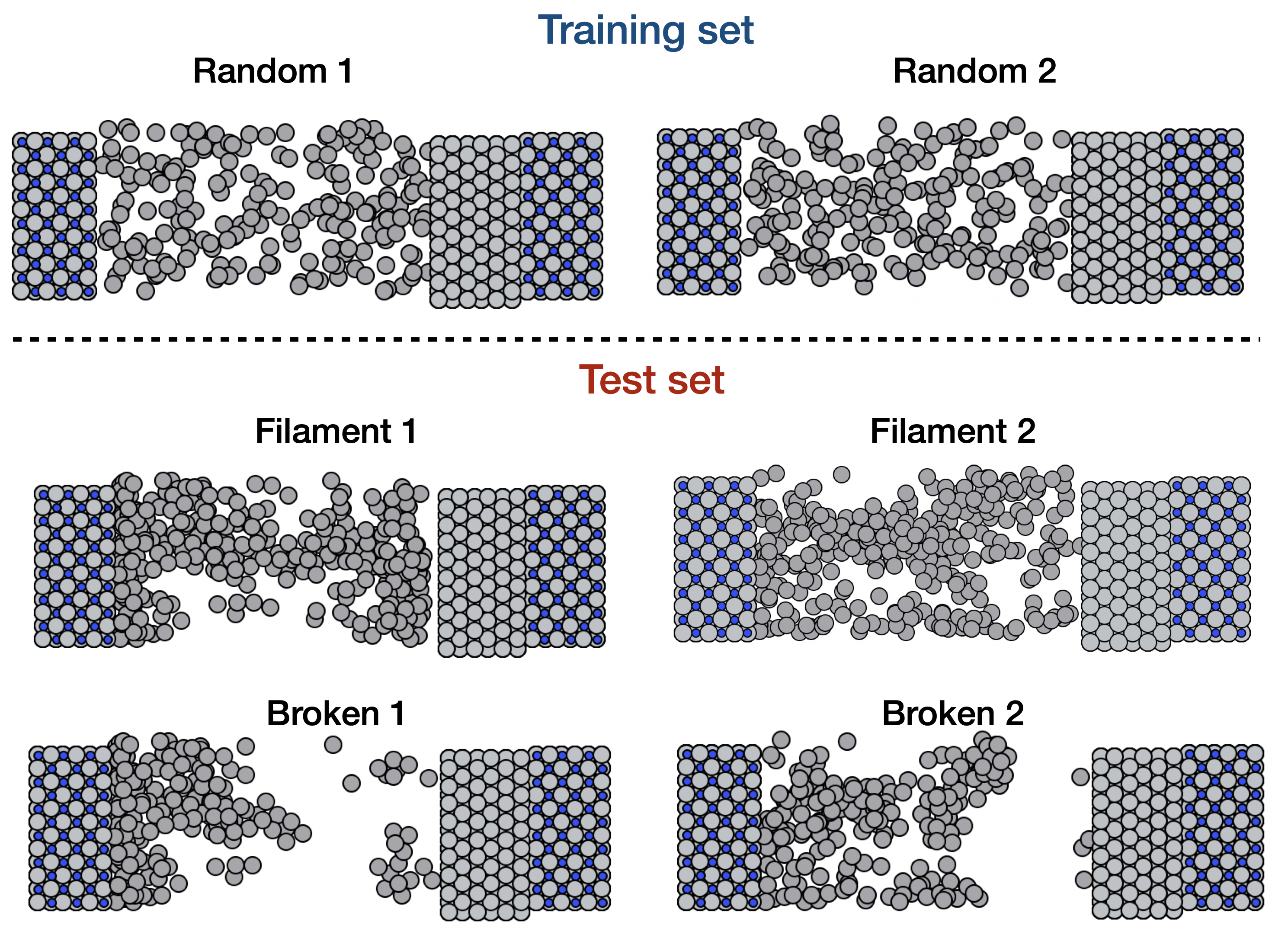}
    \caption{Training (left) and testing (right) TiN- HfO$_2$- Ti/TiN VCM structures with the Hf and O atoms omitted for better visualization. A total of 20 slices with 1 \AA \; thickness was extracted from the Random 1 and 2 devices with arbitrary vacancy distributions to train the ML model. Another unseen 1-\AA \; slice from Random 2 was used for validation. The trained model was then tested on unseen samples with either two fully-formed (Filament 1 and 2) or two broken (Broken 1 and 2) filament configurations.}
    \label{fig:structures}
\end{figure}

\begin{figure}[!t]
    \centering
    \includegraphics[width=0.8\linewidth]{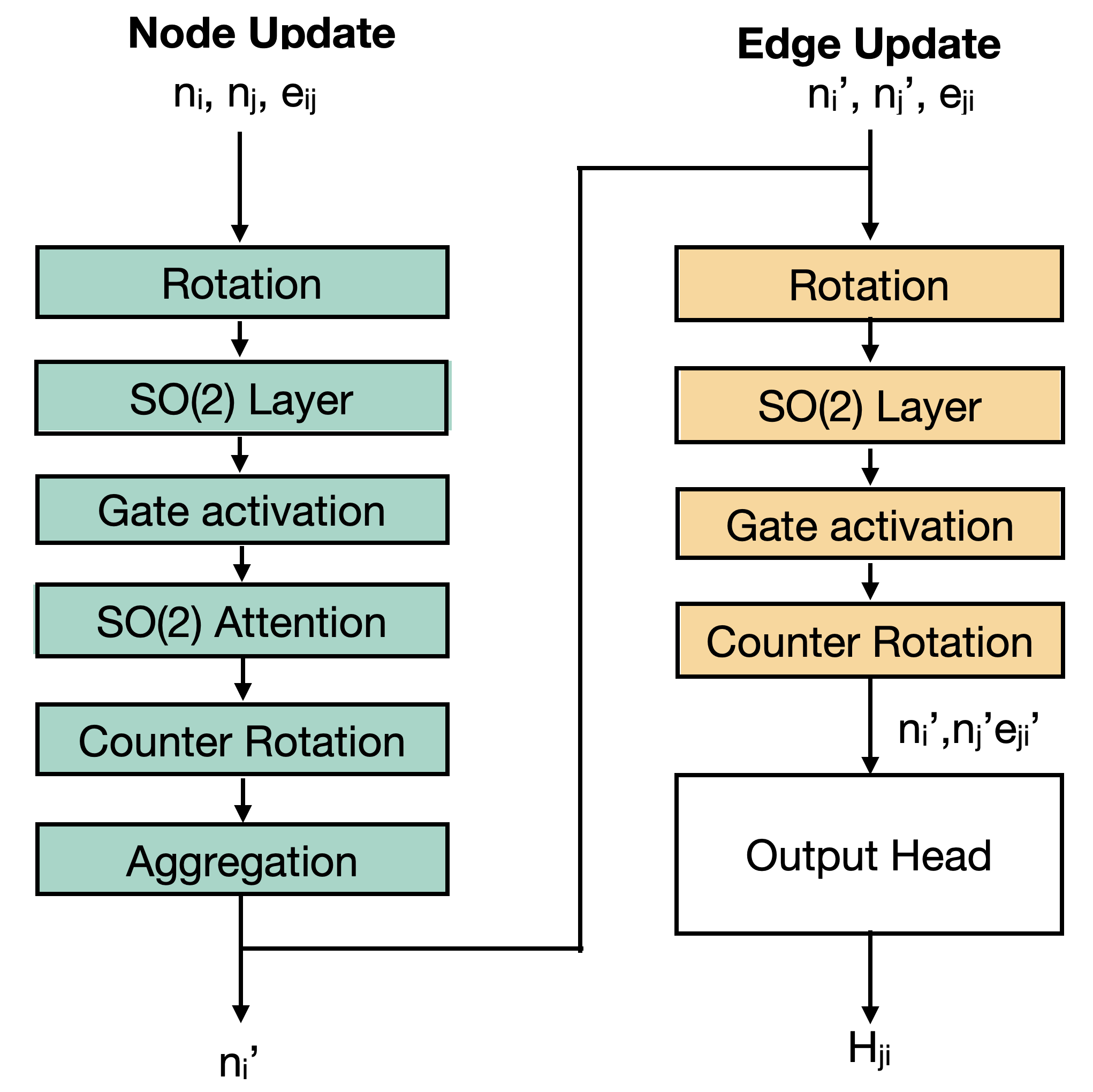}
    \caption{Higher level overview of the equivariant graph neural network architecture used in this work. It is a strictly local network consisting of a single message passing layer, with node (edge) embeddings $n_i$ ($e_{ji}$) representing atoms (interactions between atoms). The nodes aggregate messages from each other, weighted by an attention layer, with the output embeddings being used to update the edges. The updated embeddings $n_i'$ and $e_{ji}'$ are then fed into an output head to reconstruct the Hamiltonian sub-blocks. }
    \label{fig:network}
\end{figure}

\section{Method} 

\subsection{Network Architecture}
One of the key challenges in machine learning electronic structure predictions (MLESP) is the rotational covariance of the Hamiltonian matrix involved. In a localized spherical harmonic basis, a geometric rotation of the input geometry leads to a change in the Hamiltonian matrix. Equivariant graph neural networks have been proposed as a solution to drastically reduce the data required to learn such transformation \cite{TFNs}. The organization of our EGNN is presented in Fig.~\ref{fig:network}. The equivariant convolutions embed rotational covariance as a physical constraint within the network \cite{so2}. The nodes (edges) of our graph represent onsite (offsite) Hamiltonian blocks, and the outputs of the nodes are used to update the edges. To enable large-scale \textbf{H} predictions, strict locality is enforced \cite{allegro}, while multi-headed attention is included \cite{equiformerv2} to better distinguish between complex atomic environments. Finally,  the node and edge embeddings are fed into an output head to reconstruct the Hamiltonian blocks \cite{deeph3}.  

\subsection{Experiment Setup}
Our task is to predict the Hamiltonian matrix of TiN- HfO$_2$- Ti/TiN valence change memory cells with different vacancy configurations, as illustrated in Fig.~\ref{fig:structures}. The displayed snapshots were created through kinetic Monte Carlo simulations when running a ``current vs. voltage'' sweep of the devices under consideration \cite{Kaniselvan2023}. They consist of two broken and two formed filament configurations with a wide range of electrical conductivity. This quantity is highly sensitive to the distribution of oxygen vacancies within the central HfO$_2$ switching layer. 

We first train a single model on the Hamiltonian matrix of two device structures with uniformly distributed vacancies. Both $\mathbf{H}$ were produced with the CP2K DFT package \cite{cp2k}, which relies on a localized basis set of Gaussian-type orbitals. Note that the training examples were not generated via KMC, but by randomly inserting oxygen vacancies into device structures with a stoichiometric oxide layer. This makes them significantly different from the clustered, physically meaningful vacancy configurations that are part of the test set. The large discrepancies between the training and test samples is essential to rigorously assess the model's ability to learn a generalizable, useful function and apply it to unseen instances.  

\subsection{Simulation Workflow}
\begin{figure*}[!t]
    \centering
    \includegraphics[width=0.9\linewidth]{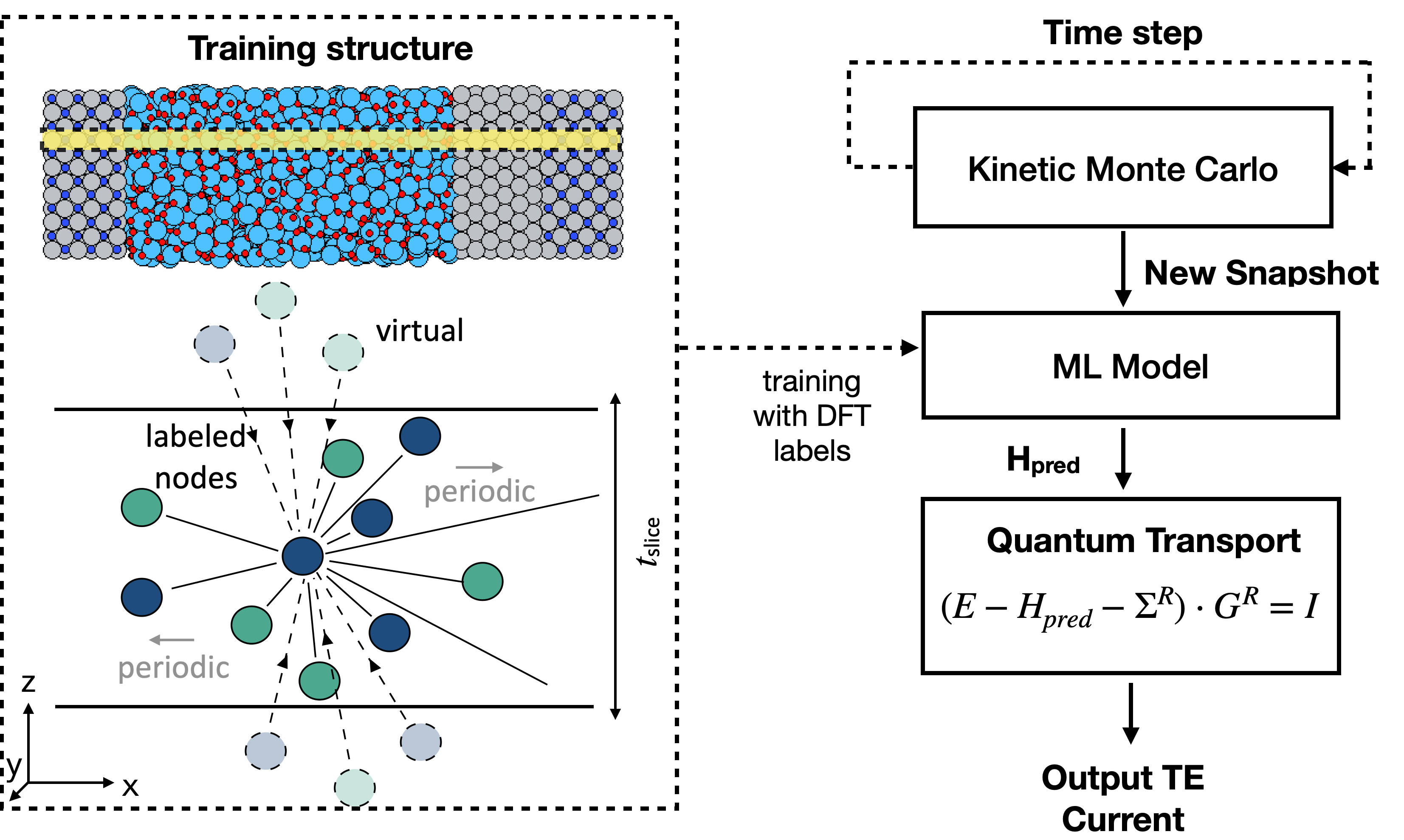}
    \caption{General overview of the workflow to train and test our machine learning model. The VCM cells considered (top left) consist of a HfO$_2$ layer (blue: Hf atoms; red: O atoms) with TiN and Ti/TiN electrodes (gray: Ti; dark blue: N). The device dimensions are set 93.4$\times$26.2$\times$26.3 \AA$^3$\; along the $x$, $y$, and $z$ directions. Slices in the $x$-$z$ plane (yellow) are used to train our EGNN (middle). A so-called augmented partitioning method (bottom left) \cite{xia2025} is leveraged to ensure that the atomic connectivity is preserved at the slice boundaries. The trained model is then used to directly infer the Hamiltonian matrix $H_{pred}$ of test structures generated with KMC at different times of a full ``I-V'' sweep. Finally, $\mathbf{H}_{pred}$ serves as input to quantum transport calculations. The transmission function $T(E)$ and electrical current $I_d$ are the final outcomes. }
    \label{fig:approach}
\end{figure*}

Our workflow is illustrated in Fig.~\ref{fig:approach}. During training, augmented partitioning \cite{xia2025} is applied to allow for large structures to be divided into multiple slices and fit into the memory of a single GPU. Importantly, the atomic connectivity to neighbor partitions is fully accounted for to maintain high prediction accuracy. Also, partitioning occurs longitudinally, in the $x-y$ plane in Fig.~\ref{fig:approach}, to capture the interface between the different materials composing the VCM stack (TiN, Ti, HfO$_2$). 

The trained EGNN is then tested on the full structure of the selected device examples by constructing/predicting their Hamiltonian matrix $\mathbf{H}$. The latter are finally inserted into our in-house quantum transport simulator \cite{omen} that returns the corresponding energy-resolved transmission function $T(E)$ and electrical current $I_d$ with the non-equilibrium Green's function (NEGF) formalism. In all cases, a voltage of 1 V is applied between both TiN electrodes of the VCM cells. The obtained results for the entries of $\mathbf{H}$, $T(E)$, and $I_d$ are compared to reference DFT calculations.   

\section{Results} 

\subsection{Hamiltonian Prediction}
\begin{figure*}[!t]
    \centering
    \includegraphics[width=1\linewidth]{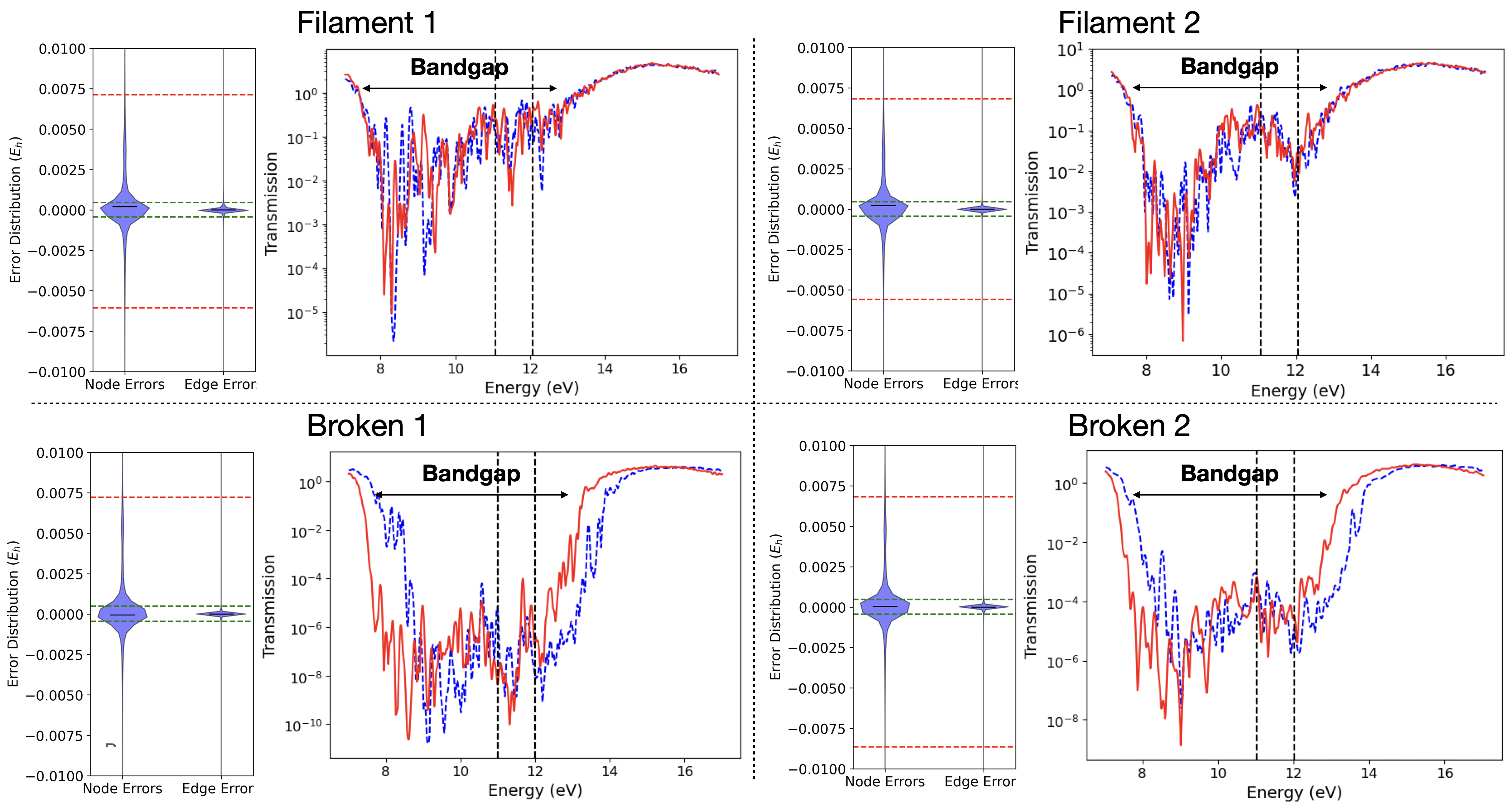}
    \caption{Prediction results for the four TiN- HfO$_2$- Ti/TiN VCM test structures from Fig.~\ref{fig:structures}. Each of them contains two sets of data. Left: Violin plot of the error distribution for the predicted Hamiltonian matrix $\mathbf{H}_{pred}$. The 5th and 95th percentile values of $\epsilon_{n}$ (node error) and $\epsilon_{e}$ (edge error) are indicated using red and green dashed lines, respectively. Note that for the "Broken 1" case, the 5th percentile lies outside of the plotted range. Right: Comparison between the energy-resolved transmission function $T(E)$ as obtained with a predicted Hamiltonian ($\mathbf{H}_{pred}$, solid red curves) with a reference DFT Hamiltonian ($\mathbf{H}_{DFT}$, dashed blue curves) for the Filament 1, Filament 2, Broken 1, and Broken 2 TiN- HfO$_2$- Ti/TiN VCM configurations from Fig.~\ref{fig:structures}. A bias of 1 V is applied between both metallic electrodes of the devices. The corresponding Fermi window at room temperature is delimited by the black dashed lines. The apparent HfO$_2$ band gap is indicated by the double arrows.}
    \label{fig:results}
\end{figure*}

The predicted node ($\epsilon_{n}$) and edge ($\epsilon_{e}$) errors of the predicted Hamiltonian matrices with respect to DFT are summarized in Table \ref{table:current} and their distributions are visualized in Fig.~\ref{fig:results} in the form of violin plots for the four TiN- HfO$_2$- Ti/TiN VCM configurations from Fig.~\ref{fig:structures}. The model performs consistently across different examples, with prediction errors per entry lying within a small range (1.54 - 1.82 $ mE_{H}$ for nodes and $\sim$0.12 $mE_{H}$ for edges), with minimal outliers in all cases. The total errors, averaged over all Hamiltonian entries, are  between 3.39 and 3.58 meV. As such, they are very close to the state-of-the-art (2.2 meV) \cite{deeph2}, but for much larger structures (5,268 vs. $\leq$150 atoms).
\begin{table}[ht]
    \centering
    \footnotesize
    \vspace{0pt}
        \caption{Summary of node ($\epsilon_{n}$) and edge ($\epsilon_{e}$) prediction errors for the different test structures considered in this work. The Mean Absolute Error (MAE) is reported in all cases. The electrical current values, as computed with a DFT ($I_{ref}$) and machine-learned ($I_{pred}$) Hamiltonian are listed as well.}
    \begin{tabular}{lrrll} \toprule
        Device& {$\mathbf{\epsilon_{n}}[mE_h]$} & {$\mathbf{\epsilon_{e}}[mE_h]$}  & $I_{ref}$ [A]&$I_{pred}$ [A]\\ \midrule
        Broken 1& 1.82& 0.12& 3.38$\times 10^{-11}$&4.35$\times 10^{-10}$\\
        Broken 1& 1.65& 0.12& 8.00$\times 10^{-9}$&4.66$\times 10^{-9}$\\
        Filament 1& 1.62& 0.12& 1.48$\times 10^{-5}$&1.28$\times 10^{-5}$\\
        Filament 2& 1.54& 0.12& 6.99$\times 10^{-6}$&4.59$\times 10^{-6}$\\
 \bottomrule
    \end{tabular}
    \label{table:current}
\end{table}

\subsection{Transmission and Current}
The predicted energy-resolved transmission functions $T(E)$ are plotted in Fig.~\ref{fig:results}. Although the underlying Hamiltonian entries do not differ by more than a few meV from their DFT reference, the corresponding $T(E)$ only qualitatively agree. The trends are the same in both cases, the band edges of the HfO$_2$ switching layers are well reproduced, but several features, especially transmission peaks, are not accurately captured. On the other hand, the predicted electrical currents, as computed from the transmission functions through the Landauer-B\"uttiker formula \cite{Brandbyge2002}, are close enough to their DFT counterparts to assess the conductance state of the device under test (see Table \ref{table:current}). Generally, it can be observed that our ML model is more accurate for configurations with fully formed filaments than for samples with broken filaments where the transmission is much smaller and therefore more sensitive to errors in the predicted Hamiltonian matrices.

Since direct ML inferences are much faster than DFT (2 seconds for the forward pass vs. 3.94 node hours for DFT), our results indicate that electronic structure predictions could facilitate the investigation of devices with evolving morphologies. By replacing DFT with ML, we can rapidly construct the Hamiltonian matrices of hundreds of intermediate samples along the high-to-low resistance transition of VCM cells, thus fully amortizing the training cost ($<$40 node hours). This will, however, require further enhancement in the accuracy of the ML models.

\section{Conclusion} We presented an ML-based approach that can be integrated into a large-scale device simulation platform capable of computing from first-principles the ``I-V'' characteristics of atomic structures changing with time. Our model  not only bypasses computationally intensive DFT calculations, it can potentially also be used to construct the Hamiltonian matrix of systems with sizes beyond current DFT capabilities. Future work includes improving the prediction accuracy of our framework by incorporating more  training data and adopting more expressive network architectures. Other applications such as phase-change memories can be envisioned. They undergo gradual amorphous-to-crystalline transitions whose electronic properties could be predicted with ML instead of being computed with DFT \cite{zhou2023}.


\bibliographystyle{unsrt}
\bibliography{biblio}

\end{document}